\documentclass[journal=apchd5,manuscript=article]{achemso}
\usepackage[T1]{fontenc}

\usepackage{amsmath}

\newcommand{\bv}[1]{\boldsymbol{#1}}
\newcommand{\tensor}[1]{\bar{\bar{#1}}}
\newcommand{\onlinecite}[1]{Ref.~[\hspace{-1 ex} \nocite{#1}\citenum{#1}]}
\DeclareMathOperator*{\diag}{diag}

\author{Gilles Rosolen}
\email{gilles.rosolen@umons.ac.be}
\affiliation[UMONS]
{Micro and Nanophotonic Materials Group, Research Institute for Materials Science and Engineering, University of Mons, Belgium}
\author{Parry Yu Chen}
\affiliation[BenGurion]
{School of Electrical and Computer Engineering, Ben-Gurion University, Israel}
\author{Bjorn Maes}
\affiliation[UMONS]
{Micro and Nanophotonic Materials Group, Research Institute for Materials Science and Engineering, University of Mons, Belgium}
\author{Yonatan Sivan}
\affiliation[BenGurion]
{School of Electrical and Computer Engineering, Ben-Gurion University, Israel}

\title{Overcoming the bottleneck for quantum computations of complex nanophotonic structures: Purcell and FRET calculations using a rigorous mode hybridization method}


\begin{document}

\begin{abstract}
A calculation of the photonic Green's tensor of a structure is at the heart of many photonic problems, but for non-trivial nanostructures, it is typically a prohibitively time-consuming task. Recently, a general normal mode expansion (GENOME) was implemented to construct the Green's tensor from eigenpermittivity modes. Here, we employ GENOME to the study the response of a cluster of nanoparticles. To this end, we use the rigorous mode hybridization theory derived earlier by D. J. Bergman [Phys. Rev. B 19, 2359 (1979)], which constructs the Green's tensor of a cluster of nanoparticles from the sole knowledge of the modes of the isolated constituent. The method is applied, for the first time, to a scatterer with a non-trivial shape (namely, a pair of elliptical wires) within a fully electrodynamic setting, and for the computation of the Purcell enhancement and F\"{o}rster Resonant Energy Transfer (FRET) rate enhancement, showing a good agreement with direct simulations. The procedure is general, trivial to implement  using standard electromagnetic software, and holds for arbitrary shapes and number of scatterers forming the cluster. Moreover, it is orders of magnitude faster than conventional direct simulations for applications requiring the spatial variation of the Green's tensor, promising a wide use in quantum technologies, free-electron light sources and heat transfer, among others.
\end{abstract}

\begin{tocentry}
\includegraphics{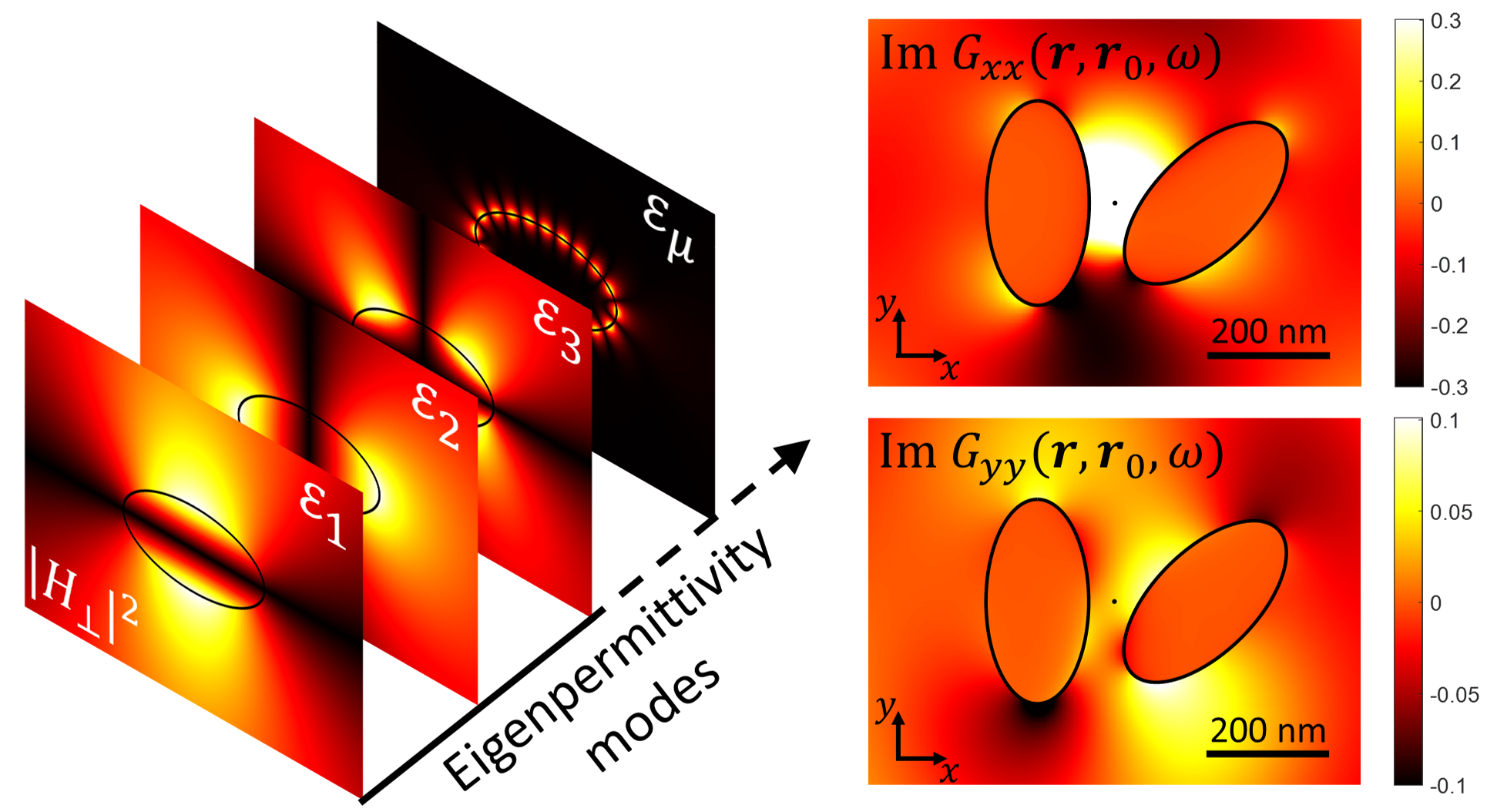}
\end{tocentry}

\section{Introduction}
It is well known that the emission properties of a quantum emitter (QE) depend on its electromagnetic (EM) environment, i.e.\ on the Green's tensor. For instance, the Purcell enhancement, which quantifies the modification of the decay rate of a QE through the EM environment~\cite{Purcell1946}, is determined by the imaginary part of the Green's tensor~\cite{Novotny2006}. Enhancing or inhibiting the Purcell enhancement has been under extensive research for its potential applications for single photon sources~\cite{Koenderink2017}, quantum sources~\cite{Dousse2010}, spectroscopy~\cite{Maslovski2019}, and so on. 

Resonant Energy Transfer (RET) is another phenomenon strongly influenced by the Green's tensor. RET is the exchange of energy between two QEs, a donor and an acceptor, and depends on the total complex Green's tensor~\cite{Novotny2006,Dung2002}. However, for a long time, it was thought that enhancing the RET comes down to enhancing the Purcell enhancement (i.e.\ considering only the imaginary part of the Green's tensor), which lead to contradictory experiments, sometimes enhancing~\cite{Andrew2000,Finlayson2001,Ghenuche2014,Ren2016}, sometimes suppressing~\cite{Reil2008,Zhao2012,Kim2012} the RET rate in the vicinity of photonic cavities or nanoparticles. Recently, it was shown that Purcell enhancement and RET are uncorrelated~\cite{Wubs2016,Cortes2018}, implying that both the real and imaginary parts of the Green's tensor need to be computed. 

Two well-known mechanisms are associated with migration of energy, depending on the distance $R$ between the QEs~\cite{Salam2018, Jones2019}: one in the near-field with a rate varying with an $R^{-6}$ dependence, a radiationless transfer called F\"{o}rster Resonant Energy Transfer (FRET), and a radiative transfer in the far-field with a rate varying with an $R^{-2}$ dependence. FRET is unique in generating fluorescence signals sensitive to molecular conformation, association, and separation and has been applied in molecular imaging techniques~\cite{JaresErijman2003}, quantum-based biosensors~\cite{Medintz2003} or in photovoltaic devices~\cite{Mor2010}. It is enhanced significantly close to nanostructures such as a graphene sheet~\cite{Biehs2013}, spherical nanoparticles~\cite{NasiriAvanaki2018}, and around a dimer of (i.e.\ two) metallic spherical particles~\cite{ZuritaSnchez2017}. Nevertheless, the challenge remains to determine the Green's tensor of complex, arbitrary, inhomogeneous, dispersive, and absorbing environments, either numerically or analytically. This problem limits the FRET rate investigations, with researchers typically resorting to effective models~\cite{Hsu2017}.

The spatial variation of the Green's tensor is known analytically for uniform media and for simple geometries. More complex structures are generally studied by direct simulation using software based on e.g.\ the finite-element method or the finite-difference time-domain method, to cite some of the well-known techniques. However, the computational cost is large, as repetitive simulations are necessary, because different polarizations and positions of the source are required. Much faster and more insightful techniques resort to an expansion of the modes of the resonator. Once the modes are known, this allows for a fast resolution of the total spatial variation of the Green's tensor, and often physical intuition can be gained e.g.\ from the study of one or two dominant modes.

For non-conservative open systems, i.e., lossy and radiative resonators, {\em eigenfrequency modes} are currently widely used, leading to the {\em quasinormal} modes (also known as resonant states)~\cite{sauvan2013theory,Muljarov_3D,Hughes_QNMs}. Eigenfrequency modal expansions were used for the study of various important physical problems like the asymmetry of the lineshape of plasmonic cavities~\cite{sauvan2013theory}, for refractive index sensing~\cite{Weiss-Giessen-prl}, for second quantization of nanocavities~\cite{franke2019quantization}, and for the derivation of scattering matrices of complex nanocavities~\cite{alpeggiani2017quasinormal}, to name a few. However, this approach is made complicated by the incompleteness of the mode set outside of the scatterers, by the need to compute them via a nonlinear eigenvalue equation, by their exponential growth in space and their non-orthogonal nature~\cite{sauvan2013theory}, by the occurrence of a large number of modes that reside in the artificial Perfectly Matched Layer etc. (see detailed discussion in~\onlinecite{Chen2019}). Intensive work of various groups has provided satisfactory solutions for these problems~\cite{Lalanne2018,muljarov2016resonant,muljarov2011brillouin,ge2014quasinormal,yan2018rigorous,weiss2017analytical}, however, correct implementation of these solutions requires significant expertise.

In contrast to the quasi-normal modes, an alternative class of {\em normal} modes that does not suffer from any of the above problems consists of {\em eigenpermittivity modes}. Indeed, in contrast to frequency eigenvalues, which are a {\em global} property of the system, the permittivity eigenvalue pertains only to a scattering element (or, an inclusion) which spans a finite portion of space. As a result,  the permittivity modes decay (rather than grow) exponentially in space, and thus, they enjoy a trivial normalization; they are orthogonal and seem to form a complete set~\cite{Bergman_Stroud_PRB_1980,Russians-permittivity-modes-book,Chen2019}. Further, they are simple to compute, since they are solutions of a linear eigenvalue problem~\cite{Chen2019}. This approach was first derived in the seventies by D. J.\ Bergman~\cite{Bergman_PRB_1979_periodic,Bergman_Stroud_PRB_1980} {\em under the quasi-static approximation}, with similar methods having been developed independently by others~\cite{Russians-permittivity-modes-book,sandu2013eigenmode}. These modes have already proven useful for the computation of effective medium parameters, bounds on them, and associated sum rules~\cite{bergman1979dielectrica,bergman1992bulk}, scattering bounds of particles~\cite{Miller2016}, as well as for the study of a wide variety of electromagnetic systems such as spasers~\cite{Bergman2003}, self-similar nanoparticle chains~\cite{li2003self}, disordered media~\cite{Stockman:2004ey}, and additional effects such as coherent control~\cite{Stockman2002} and second-harmonic generation~\cite{Li2005,Stockman:2004ey,Reddy_Sivan_SHG_BCs}, to name a few. The modes were either computed analytically for simple shapes~\cite{Bergman_PRB_1979_periodic,Bergman_Stroud_PRB_1980,Russians-permittivity-modes-book,li2003self,Farhi_Bergman_PRA_2014,Reddy_Sivan_SHG_BCs}, numerically~\cite{Bergman2003,Stockman:2004ey,Stockman2002} or asymptotically~\cite{Schnitzer_2016,Schnitzer_2019}.

Extensions of the eigenpermittivity formulation beyond the quasi-static approximation are relatively rare. The spatial variation of modes of simple shapes like a slab~\cite{Farhi_Bergman_PRA_2016}, a wire~\cite{Chen2019} and a sphere~\cite{Bergman_Stroud_PRB_1980} is known, such that the differential eigenvalue problem reduces to a complex root search problem. These can be solved using reliable algorithms (see e.g.,~\onlinecite{Parry-wire-principal_value}). More complicated scatterer systems require dedicated numerical computations~\cite{hoehne2018validation,ge2010steady}, a challenge that significantly limited the popularity of eigenpermittivity methods. This limitation was recently removed by P.Y.\ Chen \textit{et al.}\ who have applied the eigenpermittivity expansion for the computation of electromagnetic fields and the associated Green's tensor of open and lossy electromagnetic systems for general nanoparticle configurations using a commercial software (COMSOL Multiphysics)~\cite{Chen2019}; in this work, the approach was coined as a generalized normal mode expansion (GENOME). In comparison to expansions based on quasinormal modes, GENOME is trivial to implement in commercial software and converges with high accuracy.

Here, we develop another asset of GENOME, in comparison with quasinormal modes: the possibility to derive the spatial Green's tensor of a {\em cluster} of scatterers (i.e.\ an assembly of $N$ non-touching scatterers) from the knowledge of the modes of its constituents, without further mode calculations. This rigorous hybridization procedure was suggested originally by Bergman in~\onlinecite{Bergman_PRB_1979_periodic}. It may be regarded as a generalization to Maxwell's equations of the procedure employed for the calculation of molecular orbital calculations, an approach known as Linear Combination of Atomic Orbitals (LCAO)~\cite{LCAO-book,Gersten_Nitzan_review}. It was already applied for the computation of the response of a dimer (i.e.\ an assembly of two non-touching particles) from the modes of a monomer (i.e.\ one particle) under the quasi-static approximation~\cite{Bergman_PRB_1979_periodic,Nitzan_hybridization,Gersten_Nitzan_hybridization,Gersten_Nitzan_review}, as well as for the computation of the response of a periodic array of small spheres~\cite{Bergman_PRB_1979_periodic,bergman1992bulk}. Our hydridization procedure also bears similarities to multiple scattering formulations. These derive the scattering from a cluster or lattice by propagating among constituents with known scattering properties. By diagonalizing, modes can also be obtained~\cite{wang1993multiple,chen2012plane}. However, multiple scattering formulations typically use the multipole basis to express fields in the vicinity of each constituent, while our hybridisation method directly transforms constituent modes into cluster modes without any intermediate steps.

GENOME possesses a number of properties which allows this hybridisation procedure to succeed. Firstly, the modes form a discrete yet complete basis set within the inclusion, so the modes of each constituent are capable of representing any arbitrary field~\cite{Bergman_Stroud_PRB_1980}. GENOME is based on the Lippmann-Schwinger equation, which is then able to obtain the correct field everywhere from this discrete basis, even though the domain is open and infinite. This allows the multiple scattering interactions between modes of different constituents to be accounted for rigorously. Finally, since modes of the cluster and its constituents are defined by the same operator, the resulting matrix eigenvalue problem is linear. In contrast, a similar procedure would be difficult to implement for quasinormal modes, which ordinarily require a continuum of modes associated with the infinite background to ensure correct interaction between different constituents. Furthermore, quasinormal modes of the cluster and constituents have different operators, which would lead to a nonlinear matrix eigenvalue problem. 

The implementation of the rigorous hybridization method described below constitutes a rigorous generalization of the approximate quasi-static hybridization technique developed in the context of nanoplasmonics~\cite{hybridization_science_2003,hybridization_NL_2004} by Prodan, Nordlander and co-workers, while it also generalizes the previous works on eigenpermittivity modes beyond the quasi-static approximation. Our method is accurate, as validated by comparing with rigorous direct simulations, and very simple to implement. Moreover, it is faster by orders of magnitude than direct simulations for problems requiring the spatial variation of the Green's tensor, such as Purcell and FRET rate calculations, as well as for geometry optimizations. In our case, we consider a dimer of ellipsoidal rods, but the method is general for any particle shape, and any number of particles. The obtained dimer modes are used to calculate Purcell and FRET rates enhancements, for the first time to our knowledge, in the context of GENOME (and modal expansions, in general). Finally, our results confirm that the FRET rate is uncorrelated from the Purcell enhancement, making the real part of the Green's tensor essential.

Our rigorous hybridization method is derived in Section~\ref{section:hybid}, the implementation and comparison with direct simulation is successfully demonstrated in Section~\ref{section:result}, and Section~\ref{section:summary} summarizes our work and discusses several potential future steps. 

\section{Rigorous hybridization method} \label{section:hybid}

We describe the simple and efficient procedure to generate the eigenmodes of the cluster by reusing the known eigenmodes of its individual constituents. It bears many similarities to other expansion based solutions. It begins by inserting the expansion into the governing eigenvalue equation. Then, an orthonormal projection onto the constituent modes is used to produce a linear system of equations, to be solved for the solution. No additional simulation is necessary, requiring only the evaluation of overlap integrals generated during projection.

The governing eigenmode equation of the nanoparticle cluster can be expressed in integral form from the vector Helmholtz equation~\cite{Chen2019}. It may also be regarded as an eigenmode of the Lippmann-Schwinger equation for electromagnetism,
\begin{equation}
s_m\bv{E}_m(\bv{r}) = k^2 \int \tensor{G}_0(\bv{r}, \bv{r}') \theta(\bv{r}') \bv{E}_m(\bv{r}')\, d\bv{r}',
\label{eq:eigenop}
\end{equation}
where $\tensor{G}_0(\bv{r}, \bv{r}')$ is the free space Green's tensor~\cite{Novotny2006}, $k=\omega/c$ is the light wavevector, and $\bv{E}_m(\bv{r})$ is the electric field profile of the eigenmode $m$ with eigenvalue $s_m$. It is related to the eigenpermittivity $\varepsilon_m$ by $s_m = \varepsilon_b/(\varepsilon_b - \varepsilon_m)$, where $\varepsilon_b$ is the background permittivity~\cite{Chen2019}. The Heaviside type function $\theta(\bv{r})$ describes the geometry. It is zero in the background, and one in the interior of the cluster. It is a sum of disjoint geometry functions of the constituents 
\begin{equation}
\theta(\bv{r}) = \sum_a \tilde{\theta}_a(\bv{r}),
\end{equation}
with $a$ numbering the constituents, and $\tilde{\theta}_a(\bv{r})$ is non-zero only inside inclusion $a$ and zero elsewhere. By virtue of $\theta(\bv{r})$, the left hand side (LHS) of \eqref{eq:eigenop} is the field everywhere obtained from an integral defined over the interior only.

We now define the eigenmodes of each constituent; they obey an equation of identical form to that of the cluster modes defined in Eq.~\eqref{eq:eigenop},
\begin{equation}
\tilde{s}_{a,\mu}\tilde{\bv{E}}_{a,\mu}(\bv{r}) = k^2 \int \tensor{G}_0(\bv{r}, \bv{r}') \tilde{\theta}_a(\bv{r}') \tilde{\bv{E}}_{a,\mu}(\bv{r}')\, d\bv{r}',
\label{eq:eigenind}
\end{equation}
but where $\mu$ is the $\mu$th mode of constituent $a$. We have affixed tildes for quantities specific to the constituent modes.
Owing to completeness, the total interior field can be expressed as a sum of the interior fields within each constituent, permitting the expansion
\begin{equation}
\theta(\bv{r})\bv{E}_m(\bv{r}) = \sum_a\sum_\mu c_{a,\mu} \tilde{\theta}_a(\bv{r})\tilde{\bv{E}}_{a,\mu}(\bv{r}),
\label{eq:expansionint}
\end{equation}
where $c_{a,\mu}$ are weights yet to be found. These weights describe the relative contribution of the various single particle modes to the cluster mode, thus, potentially providing valueable physical insight. We restrict Eq.~\eqref{eq:eigenop} to the interior of $\theta(\bv{r})$, which allows us to insert Eq.~\eqref{eq:expansionint} to obtain
\begin{equation}
s_m \sum_a\sum_\mu c_{a,\mu}\tilde{\theta}_a(\bv{r})\tilde{\bv{E}}_{a,\mu}(\bv{r}) = k^2 \sum_a\sum_\mu c_{a,\mu} \int \tensor{G}_0(\bv{r}, \bv{r}') \tilde{\theta}_a(\bv{r}') \tilde{\bv{E}}_{a,\mu}(\bv{r}')\, d\bv{r}',
\end{equation}
which is simplified using the eigenvalue equation for the constituents \eqref{eq:eigenind} to give
\begin{equation}
s_m \sum_a\sum_\mu c_{a,\mu}\tilde{\theta}_a(\bv{r})\tilde{\bv{E}}_{a,\mu}(\bv{r}) = \sum_a\sum_\mu c_{a,\mu} s_{a,\mu} \tilde{\bv{E}}_{a,\mu}(\bv{r}).
\label{eq:projresult}
\end{equation}
This step is possible because kernels of the integrals in Eqs.~\eqref{eq:eigenop} and \eqref{eq:eigenind} are identical, corresponding to the free space Green's tensor. The only difference between the two sides of Eq.~\eqref{eq:projresult} is that the LHS is valid only in the interiors, while the right hand side is valid everywhere. Projection onto an adjoint mode $\tilde{\bv{E}}^\dagger_{b,\nu}$ then gives
\begin{align}
s_m \sum_a\sum_\mu c_{a,\mu}\int \tilde{\bv{E}}^\dagger_{b,\nu}(\bv{r}) \tilde{\theta}_b(\bv{r}) \tilde{\theta}_a(\bv{r}) \tilde{\bv{E}}_{a,\mu}(\bv{r})\, d\bv{r} &= \sum_a\sum_\mu c_{a,\mu} s_{a,\mu} \int \tilde{\bv{E}}^\dagger_{b,\nu}(\bv{r}) \tilde{\theta}_b(\bv{r}) \tilde{\bv{E}}_{a,\mu}(\bv{r})\, d\bv{r}, \nonumber \\
s_m c_{b,\nu} &= \sum_a\sum_\mu V_{b,\nu;a,\mu} \tilde{s}_{a,\mu} c_{a,\mu},
\label{eq:eigencluster}
\end{align}
where we have simplified the LHS using the orthogonality relation
\begin{equation}
\int \tilde{\bv{E}}^\dagger_{b,\nu}(\bv{r}) \tilde{\theta}_b(\bv{r}) \tilde{\theta}_a(\bv{r}) \tilde{\bv{E}}_{a,\mu}(\bv{r})\, d\bv{r} = \delta_{ab} \delta_{\mu\nu}.
\end{equation}
This states that modes belonging to different constituents are orthogonal by virtue of the disjoint $\tilde{\theta}$ functions, while the different modes belonging to the same inclusion are mutually orthogonal. In Eq.~\eqref{eq:eigencluster}, we have also defined the overlap integrals
\begin{equation}
V_{b,\nu;a,\mu} = \int \bv{E}^\dagger_{b,\nu}(\bv{r}) \tilde{\theta}_b(\bv{r}) \bv{E}_{a,\mu}(\bv{r})\, d\bv{r},
\label{eq:matrixV}
\end{equation}
which overlaps known modes of different constituents evaluated entirely within the bounds of constituent $b$. Physically speaking, the overlaps describe how a certain mode of one of the inclusions is scattered, or perturbed by the other inclusions in the system. Finally, the linear eigenvalue equation \eqref{eq:eigencluster} can be cast in matrix form to obtain the modes of the cluster,
\begin{equation}
s \bv{c} = \bv{V}\diag(\tilde{s}_{a,\mu})\bv{c},
\label{eq:eigenclustermatrix}
\end{equation}
which can be solved for the coefficients. We emphasize that the system of equations~(\ref{eq:eigenclustermatrix}) is {\em significantly smaller} than the corresponding system of equations that would arise if this problem was solved using, e.g., a finite element approach. Indeed, the size of our equation set~(\ref{eq:eigenclustermatrix}) is determined by the number of modes that have a non-negligible contribution to the solution, whereas in a direct finite-element calculation (e.g., of the vector Helmholtz equation, the differential form of Eq.~(\ref{eq:eigenop})), the number of equations that has to be solved is determined by the much higher number of mesh elements. This advantage would naturally be more significant as the number and/or size and/or acuteness of the geometrical features of scatterers in the cluster increases. This unique favourable scaling is expected to make GENOME more attractive for complex systems.

One final step exists after the coefficients $c_{a,\mu}$ in Eq.~\eqref{eq:expansionint} are found. The expansion \eqref{eq:expansionint} is only valid inside the inclusions, so to extend this to the whole domain, we insert it back into the defining eigenvalue problem \eqref{eq:eigenop},
\begin{equation}
\begin{aligned}
s_m\bv{E}_m(\bv{r}) &= k^2 \sum_a \sum_\mu c_{a,\mu} \int \tensor{G}_0(\bv{r}, \bv{r}') \tilde{\theta}_a(\bv{r}') \tilde{\bv{E}}_{a,\mu}(\bv{r}')\, d\bv{r}'\\
&= \sum_a \sum_\mu c_{a,\mu} s_{a,\mu} \tilde{\bv{E}}_{a,\mu}(\bv{r})
\label{eq:expansionext}
\end{aligned}
\end{equation}
where the result was simplified using the constituent eigenvalue problem~(\ref{eq:eigenind}). We implemented this expansion~\eqref{eq:expansionext}, which is usually preferable to the expansion \eqref{eq:expansionint} even in the interior of the inclusion, since it often converges faster. 

Finally, the Green's tensor can be obtained via GENOME,~\cite{Chen2019} using the cluster modes just found,
\begin{equation}
\begin{aligned}
\tensor{G}(\bv{r}, \bv{r}') = \tensor{G}_0(\left|\bv{r}- \bv{r}'\right|) + \frac{1}{k^2} \sum_m \frac{\varepsilon_i - \varepsilon_b}{(\varepsilon_m - \varepsilon_i)(\varepsilon_m - \varepsilon_b)} \bv{E}_m(\bv{r}) \otimes \bv{E}_m^\dagger(\bv{r}'), \label{Eq:Green}
\end{aligned}
\end{equation}
where $\varepsilon_i$ is the permittivity of the inclusion, the eigenpermittivities $\varepsilon_m = \varepsilon_b(1 - s_m^{-1})$, and the adjoint mode is
the mode itself
\begin{equation}
\bv{E}_\mu^\dagger(\bv{r}) = \bv{E}_\mu(\bv{r}).
\end{equation}
This simple adjoint holds true for all non-degenerate modes, so is applicable to all modes employed in this paper, as ellipses are not sufficiently symmetric to produce degeneracies. Two contributions are present in Eq.~\eqref{Eq:Green}. Firstly, $\tensor{G}_0(\left|\bv{r}- \bv{r}'\right|)$ is the Green's tensor of the homogenous background in the absence of any scatterers. It has a known form for the vector Helmholtz equation in 1D, 2D, and 3D. Secondly, the entire contribution due to the cluster is expressed as a sum over its eigenmodes. Its variation over source $\bv{r}'$ and detector $\bv{r}$ coordinates is separable, both expressed in terms of the spatial variation of the same set of modes. It is also in some sense diagonal, since only a single summation index $m$ exists. This simple form allows the variation of $\tensor{G}(\bv{r}, \bv{r}')$ to be mapped easily, including derivative Purcell factor and RET rates.

\section{Results} \label{section:result}

\begin{figure*}
     \centering
     \includegraphics[width=0.9\textwidth]{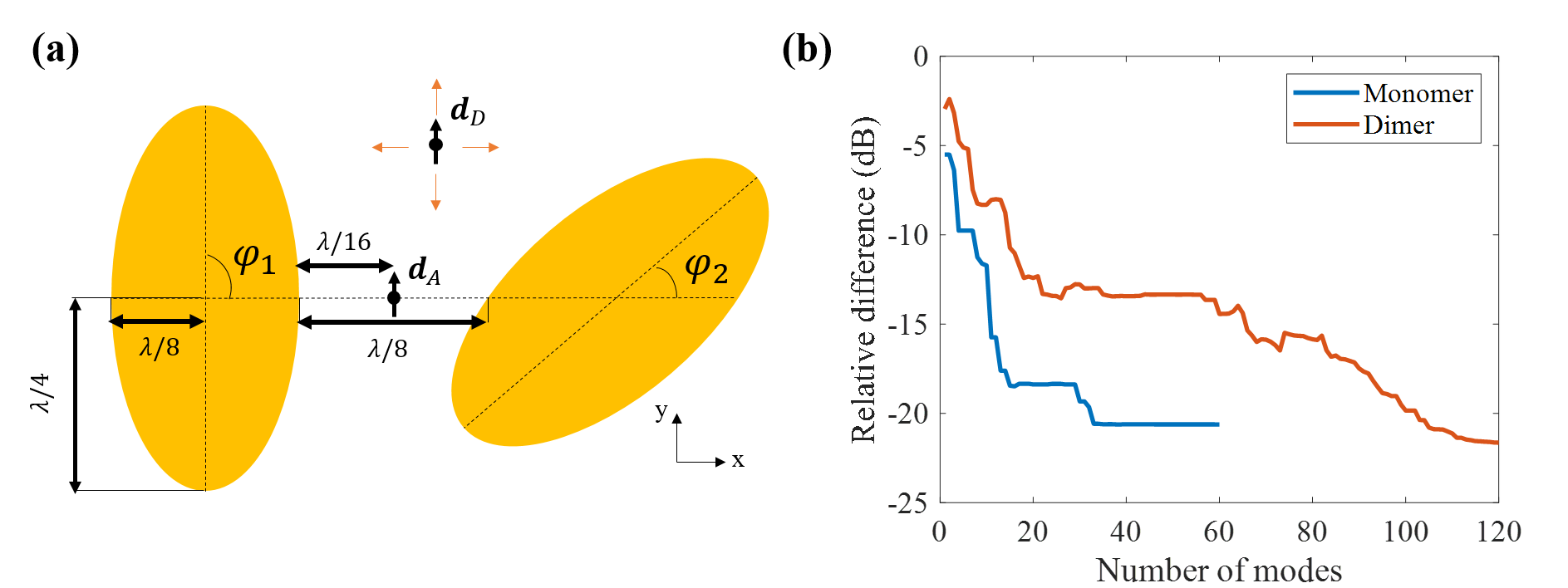}
     \caption{\textbf{Ellipse dimer considered and method convergence.} (a) The $y$-oriented dipole ($\bv{d}_A$) is at the center of the gap in-between the two ellipsoidal rods. In the case of FRET, the donor dipole ($\bv{d}_D$) varies position to generate the FRET map. The ellipse are tilted around their center by $\varphi_1=90^{\circ}$ and $\varphi_2=45^{\circ}$ such that the gap size remains $\lambda/8$. (b) $L_2$-norm of the relative difference between direct simulations (COMSOL) and our mode expansion technique (GENOME), comparing the imaginary part of the electric field following Eq.~(\ref{eq:L2norm}). The monomer case is represented for comparison, resolving the geometry without the rod on the right.} \label{fig:sheme_struct}
 \end{figure*}

\begin{figure}[t]
\center
\includegraphics[width=0.5\textwidth]{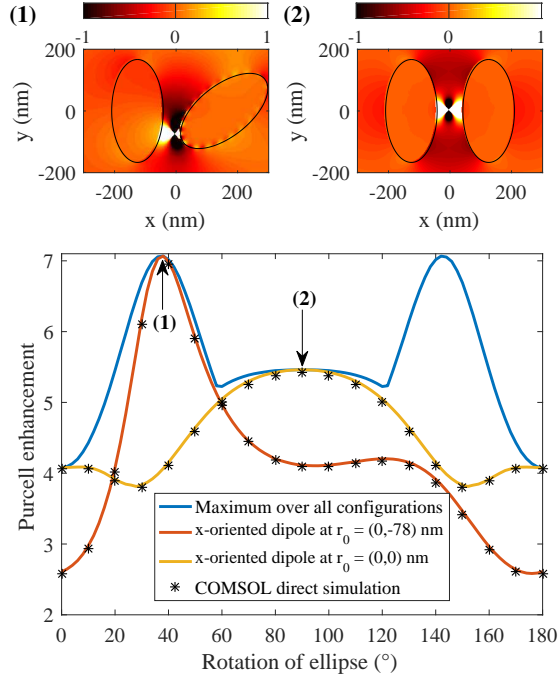}
\caption{\textbf{Maximum Purcell enhancement} over the ellipse rotation ($\varphi_2$), the dipole position along line $x=0$, and the dipole orientation. The blue line represents the maximum of the Purcell enhancement computed comparing all dipole orientations (181 points) and positions (301 points). The orange line represents the $x$-oriented dipole at $y=-78$ nm, reaching a maximum for an ellipse titled by $\varphi_2=38^\circ$. The yellow line represents the same dipole at $y=0$. The black asterisks are COMSOL direct simulations, in excellent agreement with our hybridization method. The real part of the normalized $x$-component of the electric field is plotted above for the two maxima (1) and (2).}
    \label{fig:PurcEnMax}
\end{figure}

To exemplify the method, we implement a non-trivial solution, solving the Purcell enhancement and the FRET of QEs near an ellipse dimer. The considered two-dimensional $z$-invariant structure is represented in Figure~\ref{fig:sheme_struct}a: the two ellipses have the same dimensions, with semi-axes of $\lambda/4$ and $\lambda/8$, where $\lambda$ is the wavelength. They are tilted by $\varphi_1=90^{\circ}$ and $\varphi_2=45^{\circ}$, except in the optimization discussion of Figure~\ref{fig:PurcEnMax}, where $\varphi_2$ varies. For FRET rate calculations, the donor QE is modeled by a dipole $\bv{d}_D$ and will vary in position, while the acceptor dipole $\bv{d}_A$ is fixed in the gap center, at a distance of $\lambda/16$ from the ellipses. For Purcell enhancement calculations, only a donor dipole of varying position is considered.

We use COMSOL Multiphysics to compute the eigenpermittivity modes of one elliptical rod. The software's in-build eigenfrequency mode solver is adapted with a {\em trivial} substitution trick to allow the eigenfrequencies to be reinterpreted as eigenpermittivities $\varepsilon_\mu$ (see~\onlinecite{Chen2019}); its implementation does not require any specialized knowledge as is required for the implementation of quasi-normal modes. More specifically, the rod permittivity is set to unity and the background medium of permittivity $\varepsilon_b$ is set to be dispersive, such that the wavenumber remains equal to $\sqrt{\varepsilon_b} k$ regardless of the COMSOL eigenfrequency. The system has a radius of $1.5\lambda$ and is enclosed within perfectly matched layers of $0.5\lambda$ thickness, and the mesh elements of the rod have a maximum size of $\lambda/100$. In this paper, we fix the wavelength at $\lambda = 670$nm, leading to a permittivity of the silver rod of $\varepsilon_i = -20.8 + 0.43j$.

We implement the 
rigorous hybridization formula~(\ref{eq:expansionext}) in Matlab~\cite{OnlineCodes}, importing from COMSOL the modal electric field profiles $\bv{E}_{\mu}(\bv{r})$ and their corresponding eigenpermittivities $\varepsilon_{\mu}$. We perform the integration on the mesh elements for the overlap integrals in matrix $\bv{V}$ (Eq.~\ref{eq:matrixV}), and from the solution of the eigenvalue problem (Eq.~(\ref{eq:eigenclustermatrix})), we build the new basis of hybridized modes $\bv{E}_m(\bv{r})$ as defined by Eq.~(\ref{eq:expansionext}). The Green's tensor is then computed from Eq.~(\ref{Eq:Green}). 

The convergence of the method is illustrated in Figure~\ref{fig:sheme_struct}b for metallic silver rods. We employ a pseudo-L$_2$-norm for the relative difference of the imaginary part of the electric field components:
\begin{align}
\Delta \mathrm{Im}(E) = 10\log_{10} \left[ \frac{\int \mathrm{Im}(E_x-E_{x,ref})^2+\mathrm{Im}(E_y-E_{y,ref})^2 dA}{A N}\right] \label{eq:L2norm}
\end{align}
where the reference solution is a COMSOL direct simulation i.e.\ the direct calculation of the response of the dipole $\bv{d}_A$ with the dimer, $A$ is the area of an integration domain including the dimer (in the case of Figure~\ref{fig:sheme_struct}b, $\lambda = 670$ nm, the area is $800 \times 600$ nm and centered on the dipole position $\bv{d}_A$) and $N$ is a normalization factor taken to be the square of the maximal value of the imaginary part of the field of the COMSOL solution. Note that $E_z = 0$ since we solve for the TM polarization.

In Figure~\ref{fig:sheme_struct}b, good convergence is obtained after 20 modes for the elliptical monomer and saturates at -20dB. This is due to numerical errors in the field discretization of the COMSOL direct simulation taken as the reference simulation, and in the COMSOL mode generation constituting the building blocks of our Green's tensor. Moreover, we limit the mode search to 60 modes, truncating from the basis higher order radial modes that would further enhance the accuracy. The constructed basis of the ellipse dimer expands on 120 modes, twice the number of modes of the monomer case, and Figure~\ref{fig:sheme_struct}b shows a good convergence (-20 dB) after 90 modes. The number of modes required to obtain convergence, and therefore the computation time, depends on the permittivity of the rod, inclination of the rod, the distance between the rods and the position of the dipole (not shown here). 

\begin{figure}[t]
\center
\includegraphics[width=1.0\textwidth]{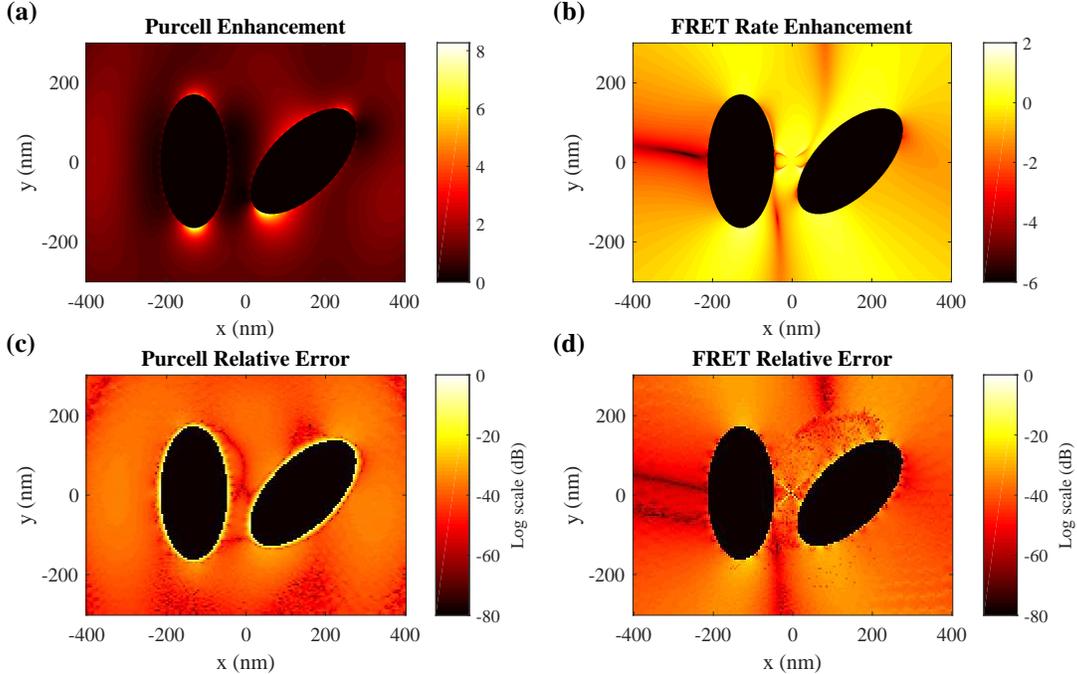}
\caption{\textbf{Efficient mapping of Purcell enhancement and FRET rates.} The geometry of Figure \ref{fig:sheme_struct}a is resolved for \textbf{(a)} the Purcell enhancement and \textbf{(b)} the FRET rate enhancement. L$_2$-norm of the relative difference with COMSOL simulations is represented for \textbf{(c)} the Purcell enhancement and \textbf{(d)} the FRET rate enhancement. This degree of convergence (mean error is -40 dB for both maps) is obtained for 50 monomer modes. The map generation takes 11 minutes for the modes calculation and an additional 65 seconds for the hybridization method (with room for improvement), while COMSOL takes 17 hours.} \label{fig:FandP}
\end{figure}

The modes obtained from the hybridization formulation are now applied to compute the Purcell enhancement (Figure~\ref{fig:FandP}a) and the FRET rate enhancement (Figure~\ref{fig:FandP}b). For direct simulations, we loop over the positions of the dipole $\bv{d}_D$, and we monitor the flux through a small box surrounding the dipole to compute the Purcell enhancement, while measuring the field at the position of $\bv{d}_A$ for the FRET rate. Although one direct simulation is fast, repeated simulation of varying emitter positions is time-consuming and can be prohibitive. With GENOME, the spatial dependence of the Green's tensor is constructed extremely rapidly from the knowledge of the spatial profile of the modes of the monomer, leading straightforwardly to the maps in Figure~\ref{fig:FandP}. For comparison, these maps take 12 minutes to compute with GENOME on a desktop computer: 11 minutes for the modes and 65 seconds to construct the matrix $\bv{V}$, solve the eigenvalue problem of the dimer structure and retrieve the Purcell and FRET rate enhancement maps. Meanwhile, direct COMSOL simulations take 17 hours, even though the mesh is coarser than for the GENOME mode calculation: the maximum element size within the rod is $\lambda/60$ for COMSOL and $\lambda/100$ for the mode search. Moreover, GENOME also gives access to the two other polarization directions in 2 seconds, while COMSOL needs another 17 hours for each polarization. Further speed improvements can be realized by exploiting symmetries in the matrix $\bv{V}$ and with an implementation method that reduces the number of connections between COMSOL and Matlab. Note that maps of FRET enhancement computed via direct simulations can be fast when one emitter (donor or acceptor) is fixed, thanks to the property $G_{ij}(\bv{r}_D,\bv{r}_A,\omega) = G_{ji}(\bv{r}_A,\bv{r}_D,\omega)$ of the Green's tensor. For results where neither emitter is fixed, GENOME is faster.

The Purcell enhancement $\Gamma_e$ is given by
\begin{align}
\Gamma_e &=  \frac{ \bv{d}_D^* \cdot \mathrm{Im} \left( \tensor{G}(\bv{r}_D,\bv{r}_D,\omega_0)   \right) \cdot  \bv{d}_D }{ \bv{d}_D^* \cdot  \mathrm{Im} \left( \tensor{G}_0(\bv{r}_D,\bv{r}_D,\omega_0)\right) \cdot \bv{d}_D}. \label{eq:PurcellEn} 
\end{align}
It is represented in Figure~\ref{fig:FandP}a, for all positions of a $y$-polarized dipole outside the dimer, with the interior of the dimer represented in black. The strongest Purcell enhancement, exceeding $\Gamma_e=6$, is observed for a dipole close to the tips of the ellipses. In Figure~\ref{fig:FandP}c, we compare the GENOME formulation to COMSOL direct simulations: we use the metric defined in Eq.~(\ref{eq:L2norm}), where we replace the electric field by the Purcell enhancement or the FRET rate. In general, a great accuracy of -40dB is obtained, except for a small region of 5 nm around the edges of the ellipses, where the accuracy is above -20dB. This indicates missing higher-order ultra-confined plasmonic modes in the expansion. This inaccuracy can be addressed by considering more modes in the expansion, if it is necessary for the region of interest. Note, however, that at those distances quantum corrections may be required: the point-dipole approximation is no longer valid for the QE~\cite{Rivera2016} and non-local effects cannot be neglected very close to the particle~\cite{Carminati_nonlocal_local}.

The FRET rate enhancement is computed for the same dimer in Figure~\ref{fig:FandP}b for all positions of the $y$-polarized donor outside the dimer, where the acceptor is in the center of the gap. The FRET rate $\Gamma_{DA}$ is given by~\cite{Dung2002,Wubs2016}:
\begin{align}
\Gamma_{DA}= \frac{2\pi}{\hbar^2} \left( \frac{\omega_0^2}{\varepsilon_0 c^2} \right)^2 \left|\bv{d}_A^* \cdot \tensor{G}(\bv{r}_A,\bv{r}_D,\omega_0) \cdot \bv{d}_D \right|^2 \label{eq:FRET}
\end{align}
where $\hbar$ is the reduced Planck constant, $\omega_0$ the angular frequency, $\varepsilon_0$ the vacuum permittivity, $c$ the speed of light, and where we suppose that the donor and acceptor absorption spectra in free-space are identical and are equal to a delta function centered on $\omega_0$. The FRET rate enhancement is $\Gamma_{DA}$ normalized by the FRET rate in vacuum ($\tensor{G}$ is replaced by $\tensor{G}_0$ in Equation~(\ref{eq:FRET})). The strongest enhancements appear when the donor is close to the acceptor, but also around the tips of the ellipses. On the contrary, the FRET rate is inhibited for a donor positioned on the left of the dimer. Comparing the Purcell enhancement and the FRET rate (Figures~\ref{fig:FandP}a and b) shows that the FRET rate and the imaginary part of the Green's tensor (i.e.\ Purcell, see Eq.~\ref{eq:PurcellEn}) are uncorrelated, as demonstrated in~\onlinecite{Cortes2018}, closing the on-going debate on the influence of the imaginary part of the Green's tensor (related to the local density of states) at the position of the donor on the FRET rate. The comparison with direct simulations is represented in Figure~\ref{fig:FandP}d, using a L$_2$-norm similar to Equation~(\ref{eq:L2norm}). In general, we reach an excellent agreement of -40dB, with particular zones around the edges of the ellipses amounting to a larger error of -20dB, which we relate to the missing higher-order plasmonic modes in the expansion.

Our method is powerful for applications relying on the spatial variation of the Green's tensor. Indeed, once the new mode basis of the hybridized mode is known, $\tensor{G}(\bv{r},\bv{r}',\omega_0)$ is determined over the entire space, hence allowing for diverse optimizations. In Figure~\ref{fig:PurcEnMax}, we optimize the Purcell enhancement as a function of the ellipse rotation ($\varphi_1 = 90^\circ$ and $\varphi_2$ varies), dipole orientation, and dipole location along the vertical line $x=0$. Note that the ellipse is rotated such that the gap size remains $\lambda/8$ at $y=0$. The blue line shows an optimum for an ellipse tilted by $\varphi_2=38^\circ$ and an $x$-oriented dipole at position $y=-78$ nm. The real part of the corresponding electric field ($x$-component) is represented in insert (1), showing that the maximum Purcell enhancement is obtained for a geometry where the dipole is the closest to the right rod. A local maximum of 5.5 is reached for parallel ellipses (tilted by $\varphi_1=\varphi_2=90^\circ$) and an $x$-oriented dipole at position $y=0$ nm, with the field represented in (2). The orange and yellow curves are plotted for specific dipole orientation and locations, showing excellent agreement with COMSOL direct simulations (black asterisks). The maximum appearing at $\varphi_2=142^\circ$ is due to the symmetric situation to the orange curve, with an $x$-oriented dipole at position $y=78$ nm. In this particular example, $5\times 10^6$ configurations (91 ellipse rotations, 301 dipole positions and 181 orientations) were computed in 19 minutes (with the modes already known), i.e.\ 1000 times faster than direct simulations (3 seconds per configuration).

\section{Summary} \label{section:summary}

In this paper, we implement an electrodynamic formulation of eigenpermittivity mode hybridization, retrieving the Green's tensor of a dimer of scatterers from the eigenpermittivity modes of the constituent scatterers. The formulation is successfully tested for an ellipse dimer, and the implementation for more scatterers of general and non-identical shape is straightforward. As we discussed, these calculations generalize previous works beyond the quasi-static approximation~\cite{Bergman_PRB_1979_periodic,Gersten_Nitzan_review}; they also constitute a rigorous generalization of the approximate hybridization technique developed independently later by Prodan, Nordlander and co-workers (e.g.,~\onlinecite{hybridization_science_2003,hybridization_NL_2004}), which applied a microscopic, physically-intuitive model of a quasi-free electron gas to derive the complex eigenfrequencies of various complicated plasmonic nanostructures. Indeed, our approach does not rely on assumptions such as the negligibility of the valence band electrons, incompressibility of the electron cloud, the absence of Drude losses and of interband transitions, or the jellium approximation of the ion background; in contrast, our approach accounts for absorption losses, in addition to and independently of radiation losses, thus, capturing the open and lossy nature of these systems. Finally, our calculations can be used to validate various analytic approaches already applied to the problem of particle dimers~\cite{Alex_hybridization} or even more complicated structures (e.g., in the context of Fano resonances and plasmon-induced transparency~\cite{Giannini2011,Zhang2008,Giessen_Liu_oligomers}). 

The method is in good agreement with direct simulations, and is orders of magnitude faster, even including the time required to find modes of a single particle. It is therefore suitable for applications requiring the spatial variation of the Green's tensor, as we show for maps of Purcell enhancement and FRET rates. We also envision it to play a major role in quantum computation, particularly in quantum plasmonics, to quantify the high-order decay rates of quantum emitters~\cite{Rivera2016}, for free-electron high-harmonic light sources~\cite{Rivera2019,Rosolen2018}, heat transfer and thermal emission engineering~\cite{Guo_Thermal_MTM,Guo_Thermal_MTM_TPV,Francoeur2009,Czapla2017}, van der Waals forces~\cite{Luo2014} and for disordered media~\cite{Mosk_review_NP}. For simple systems, the method also allows for a greater physical insight into the interactions among the constituents as it is a weighted sum of constituent modes of the single particle. Further work is underway to generalize the method to periodic structures, to 3D particles, to non-uniform permittivities and to anisotropic media, to ensure that GENOME benefits a wide range of research fields.

\begin{acknowledgement}
G.R.\ was supported by the Fonds De La Recherche Scientifique - FNRS. P.Y.C.\ and Y.S.\ acknowledge support by the Israel Science Foundation (ISF) grant (899/16).
\end{acknowledgement}

\bibliography{bibli}

\providecommand{\latin}[1]{#1}
\makeatletter
\providecommand{\doi}
  {\begingroup\let\do\@makeother\dospecials
  \catcode`\{=1 \catcode`\}=2 \doi@aux}
\providecommand{\doi@aux}[1]{\endgroup\texttt{#1}}
\makeatother
\providecommand*\mcitethebibliography{\thebibliography}
\csname @ifundefined\endcsname{endmcitethebibliography}
  {\let\endmcitethebibliography\endthebibliography}{}
\begin{mcitethebibliography}{81}
\providecommand*\natexlab[1]{#1}
\providecommand*\mciteSetBstSublistMode[1]{}
\providecommand*\mciteSetBstMaxWidthForm[2]{}
\providecommand*\mciteBstWouldAddEndPuncttrue
  {\def\EndOfBibitem{\unskip.}}
\providecommand*\mciteBstWouldAddEndPunctfalse
  {\let\EndOfBibitem\relax}
\providecommand*\mciteSetBstMidEndSepPunct[3]{}
\providecommand*\mciteSetBstSublistLabelBeginEnd[3]{}
\providecommand*\EndOfBibitem{}
\mciteSetBstSublistMode{f}
\mciteSetBstMaxWidthForm{subitem}{(\alph{mcitesubitemcount})}
\mciteSetBstSublistLabelBeginEnd
  {\mcitemaxwidthsubitemform\space}
  {\relax}
  {\relax}

\bibitem[Purcell(1946)]{Purcell1946}
Purcell,~E.~M. Spontaneous Emission Probabilities at Radio Frequencies.
  \emph{Physical Review} \textbf{1946}, \emph{69}, 681\relax
\mciteBstWouldAddEndPuncttrue
\mciteSetBstMidEndSepPunct{\mcitedefaultmidpunct}
{\mcitedefaultendpunct}{\mcitedefaultseppunct}\relax
\EndOfBibitem
\bibitem[Novotny and Hecht(2006)Novotny, and Hecht]{Novotny2006}
Novotny,~L.; Hecht,~B. In \emph{Principles of Nano-Optics}; Press,~C.~U., Ed.;
  Cambridge University Press, 2006\relax
\mciteBstWouldAddEndPuncttrue
\mciteSetBstMidEndSepPunct{\mcitedefaultmidpunct}
{\mcitedefaultendpunct}{\mcitedefaultseppunct}\relax
\EndOfBibitem
\bibitem[Koenderink(2017)]{Koenderink2017}
Koenderink,~A.~F. Single-Photon Nanoantennas. \textbf{2017}, \emph{4},
  710--722\relax
\mciteBstWouldAddEndPuncttrue
\mciteSetBstMidEndSepPunct{\mcitedefaultmidpunct}
{\mcitedefaultendpunct}{\mcitedefaultseppunct}\relax
\EndOfBibitem
\bibitem[Dousse \latin{et~al.}(2010)Dousse, Suffczy{\'{n}}ski, Beveratos,
  Krebs, Lema{\^{\i}}tre, Sagnes, Bloch, Voisin, and Senellart]{Dousse2010}
Dousse,~A.; Suffczy{\'{n}}ski,~J.; Beveratos,~A.; Krebs,~O.;
  Lema{\^{\i}}tre,~A.; Sagnes,~I.; Bloch,~J.; Voisin,~P.; Senellart,~P.
  Ultrabright source of entangled photon pairs. \emph{Nature} \textbf{2010},
  \emph{466}, 217--220\relax
\mciteBstWouldAddEndPuncttrue
\mciteSetBstMidEndSepPunct{\mcitedefaultmidpunct}
{\mcitedefaultendpunct}{\mcitedefaultseppunct}\relax
\EndOfBibitem
\bibitem[Maslovski and Simovski(2019)Maslovski, and Simovski]{Maslovski2019}
Maslovski,~S.~I.; Simovski,~C.~R. Purcell factor and local intensity
  enhancement in surface-enhanced {R}aman scattering. \emph{Nanophotonics}
  \textbf{2019}, \emph{8}, 429--434\relax
\mciteBstWouldAddEndPuncttrue
\mciteSetBstMidEndSepPunct{\mcitedefaultmidpunct}
{\mcitedefaultendpunct}{\mcitedefaultseppunct}\relax
\EndOfBibitem
\bibitem[Dung \latin{et~al.}(2002)Dung, Kn\"{o}ll, and Welsch]{Dung2002}
Dung,~H.~T.; Kn\"{o}ll,~L.; Welsch,~D.-G. Intermolecular energy transfer in the
  presence of dispersing and absorbing media. \emph{Physical Review A}
  \textbf{2002}, \emph{65}\relax
\mciteBstWouldAddEndPuncttrue
\mciteSetBstMidEndSepPunct{\mcitedefaultmidpunct}
{\mcitedefaultendpunct}{\mcitedefaultseppunct}\relax
\EndOfBibitem
\bibitem[Andrew(2000)]{Andrew2000}
Andrew,~P. F\"{o}rster Energy Transfer in an Optical Microcavity.
  \emph{Science} \textbf{2000}, \emph{290}, 785--788\relax
\mciteBstWouldAddEndPuncttrue
\mciteSetBstMidEndSepPunct{\mcitedefaultmidpunct}
{\mcitedefaultendpunct}{\mcitedefaultseppunct}\relax
\EndOfBibitem
\bibitem[Finlayson \latin{et~al.}(2001)Finlayson, Ginger, and
  Greenham]{Finlayson2001}
Finlayson,~C.~E.; Ginger,~D.~S.; Greenham,~N.~C. Enhanced {F}\"{o}rster energy
  transfer in organic/inorganic bilayer optical microcavities. \emph{Chemical
  Physics Letters} \textbf{2001}, \emph{338}, 83--87\relax
\mciteBstWouldAddEndPuncttrue
\mciteSetBstMidEndSepPunct{\mcitedefaultmidpunct}
{\mcitedefaultendpunct}{\mcitedefaultseppunct}\relax
\EndOfBibitem
\bibitem[Ghenuche \latin{et~al.}(2014)Ghenuche, de~Torres, Moparthi, Grigoriev,
  and Wenger]{Ghenuche2014}
Ghenuche,~P.; de~Torres,~J.; Moparthi,~S.~B.; Grigoriev,~V.; Wenger,~J.
  Nanophotonic Enhancement of the {F}\"{o}rster Resonance Energy-Transfer Rate
  with Single Nanoapertures. \emph{Nano Letters} \textbf{2014}, \emph{14},
  4707--4714\relax
\mciteBstWouldAddEndPuncttrue
\mciteSetBstMidEndSepPunct{\mcitedefaultmidpunct}
{\mcitedefaultendpunct}{\mcitedefaultseppunct}\relax
\EndOfBibitem
\bibitem[Ren \latin{et~al.}(2016)Ren, Wu, Yang, and Zhang]{Ren2016}
Ren,~J.; Wu,~T.; Yang,~B.; Zhang,~X. Simultaneously giant enhancement of
  {F}\"{o}rster resonance energy transfer rate and efficiency based on
  plasmonic excitations. \emph{Physical Review B} \textbf{2016},
  \emph{94}\relax
\mciteBstWouldAddEndPuncttrue
\mciteSetBstMidEndSepPunct{\mcitedefaultmidpunct}
{\mcitedefaultendpunct}{\mcitedefaultseppunct}\relax
\EndOfBibitem
\bibitem[Reil \latin{et~al.}(2008)Reil, Hohenester, Krenn, and
  Leitner]{Reil2008}
Reil,~F.; Hohenester,~U.; Krenn,~J.~R.; Leitner,~A. {F}\"{o}rster-Type Resonant
  Energy Transfer Influenced by Metal Nanoparticles. \emph{Nano Letters}
  \textbf{2008}, \emph{8}, 4128--4133\relax
\mciteBstWouldAddEndPuncttrue
\mciteSetBstMidEndSepPunct{\mcitedefaultmidpunct}
{\mcitedefaultendpunct}{\mcitedefaultseppunct}\relax
\EndOfBibitem
\bibitem[Zhao \latin{et~al.}(2012)Zhao, Ming, Shao, Chen, and Wang]{Zhao2012}
Zhao,~L.; Ming,~T.; Shao,~L.; Chen,~H.; Wang,~J. Plasmon-Controlled
  {F}\"{o}rster Resonance Energy Transfer. \emph{The Journal of Physical
  Chemistry C} \textbf{2012}, \emph{116}, 8287--8296\relax
\mciteBstWouldAddEndPuncttrue
\mciteSetBstMidEndSepPunct{\mcitedefaultmidpunct}
{\mcitedefaultendpunct}{\mcitedefaultseppunct}\relax
\EndOfBibitem
\bibitem[Kim \latin{et~al.}(2012)Kim, Kim, Kim, Laquai, Arifin, Lee, Yoo, and
  Sohn]{Kim2012}
Kim,~K.-S.; Kim,~J.-H.; Kim,~H.; Laquai,~F.; Arifin,~E.; Lee,~J.-K.;
  Yoo,~S.~I.; Sohn,~B.-H. Switching Off {FRET} in the Hybrid Assemblies of
  Diblock Copolymer Micelles, Quantum Dots, and Dyes by Plasmonic
  Nanoparticles. \emph{{ACS} Nano} \textbf{2012}, \emph{6}, 5051--5059\relax
\mciteBstWouldAddEndPuncttrue
\mciteSetBstMidEndSepPunct{\mcitedefaultmidpunct}
{\mcitedefaultendpunct}{\mcitedefaultseppunct}\relax
\EndOfBibitem
\bibitem[Wubs and Vos(2016)Wubs, and Vos]{Wubs2016}
Wubs,~M.; Vos,~W.~L. F\"{o}rster resonance energy transfer rate in any
  dielectric nanophotonic medium with weak dispersion. \emph{New Journal of
  Physics} \textbf{2016}, \emph{18}, 053037\relax
\mciteBstWouldAddEndPuncttrue
\mciteSetBstMidEndSepPunct{\mcitedefaultmidpunct}
{\mcitedefaultendpunct}{\mcitedefaultseppunct}\relax
\EndOfBibitem
\bibitem[Cortes and Jacob(2018)Cortes, and Jacob]{Cortes2018}
Cortes,~C.~L.; Jacob,~Z. Fundamental figures of merit for engineering
  {F}\"{o}rster resonance energy transfer. \emph{Optics Express} \textbf{2018},
  \emph{26}, 19371\relax
\mciteBstWouldAddEndPuncttrue
\mciteSetBstMidEndSepPunct{\mcitedefaultmidpunct}
{\mcitedefaultendpunct}{\mcitedefaultseppunct}\relax
\EndOfBibitem
\bibitem[Salam(2018)]{Salam2018}
Salam,~A. The Unified Theory of Resonance Energy Transfer According to
  Molecular Quantum Electrodynamics. \emph{Atoms} \textbf{2018}, \emph{6},
  56\relax
\mciteBstWouldAddEndPuncttrue
\mciteSetBstMidEndSepPunct{\mcitedefaultmidpunct}
{\mcitedefaultendpunct}{\mcitedefaultseppunct}\relax
\EndOfBibitem
\bibitem[Jones and Bradshaw(2019)Jones, and Bradshaw]{Jones2019}
Jones,~G.~A.; Bradshaw,~D.~S. Resonance Energy Transfer: From Fundamental
  Theory to Recent Applications. \emph{Frontiers in Physics} \textbf{2019},
  \emph{7}\relax
\mciteBstWouldAddEndPuncttrue
\mciteSetBstMidEndSepPunct{\mcitedefaultmidpunct}
{\mcitedefaultendpunct}{\mcitedefaultseppunct}\relax
\EndOfBibitem
\bibitem[Jares-Erijman and Jovin(2003)Jares-Erijman, and
  Jovin]{JaresErijman2003}
Jares-Erijman,~E.~A.; Jovin,~T.~M. {FRET} imaging. \emph{Nature Biotechnology}
  \textbf{2003}, \emph{21}, 1387--1395\relax
\mciteBstWouldAddEndPuncttrue
\mciteSetBstMidEndSepPunct{\mcitedefaultmidpunct}
{\mcitedefaultendpunct}{\mcitedefaultseppunct}\relax
\EndOfBibitem
\bibitem[Medintz \latin{et~al.}(2003)Medintz, Clapp, Mattoussi, Goldman,
  Fisher, and Mauro]{Medintz2003}
Medintz,~I.~L.; Clapp,~A.~R.; Mattoussi,~H.; Goldman,~E.~R.; Fisher,~B.;
  Mauro,~J.~M. Self-assembled nanoscale biosensors based on quantum dot {FRET}
  donors. \emph{Nature Materials} \textbf{2003}, \emph{2}, 630--638\relax
\mciteBstWouldAddEndPuncttrue
\mciteSetBstMidEndSepPunct{\mcitedefaultmidpunct}
{\mcitedefaultendpunct}{\mcitedefaultseppunct}\relax
\EndOfBibitem
\bibitem[Mor \latin{et~al.}(2010)Mor, Basham, Paulose, Kim, Varghese, Vaish,
  Yoriya, and Grimes]{Mor2010}
Mor,~G.~K.; Basham,~J.; Paulose,~M.; Kim,~S.; Varghese,~O.~K.; Vaish,~A.;
  Yoriya,~S.; Grimes,~C.~A. High-Efficiency {F}\"{o}rster Resonance Energy
  Transfer in Solid-State Dye Sensitized Solar Cells. \emph{Nano Letters}
  \textbf{2010}, \emph{10}, 2387--2394\relax
\mciteBstWouldAddEndPuncttrue
\mciteSetBstMidEndSepPunct{\mcitedefaultmidpunct}
{\mcitedefaultendpunct}{\mcitedefaultseppunct}\relax
\EndOfBibitem
\bibitem[Biehs and Agarwal(2013)Biehs, and Agarwal]{Biehs2013}
Biehs,~S.-A.; Agarwal,~G.~S. Large enhancement of {F}\"{o}rster resonance
  energy transfer on graphene platforms. \emph{Applied Physics Letters}
  \textbf{2013}, \emph{103}, 243112\relax
\mciteBstWouldAddEndPuncttrue
\mciteSetBstMidEndSepPunct{\mcitedefaultmidpunct}
{\mcitedefaultendpunct}{\mcitedefaultseppunct}\relax
\EndOfBibitem
\bibitem[Avanaki \latin{et~al.}(2018)Avanaki, Ding, and
  Schatz]{NasiriAvanaki2018}
Avanaki,~K.~N.; Ding,~W.; Schatz,~G.~C. Resonance Energy Transfer in Arbitrary
  Media: Beyond the Point Dipole Approximation. \emph{The Journal of Physical
  Chemistry C} \textbf{2018}, \emph{122}, 29445--29456\relax
\mciteBstWouldAddEndPuncttrue
\mciteSetBstMidEndSepPunct{\mcitedefaultmidpunct}
{\mcitedefaultendpunct}{\mcitedefaultseppunct}\relax
\EndOfBibitem
\bibitem[Zurita-S{\'{a}}nchez and
  M{\'{e}}ndez-Villanueva(2017)Zurita-S{\'{a}}nchez, and
  M{\'{e}}ndez-Villanueva]{ZuritaSnchez2017}
Zurita-S{\'{a}}nchez,~J.~R.; M{\'{e}}ndez-Villanueva,~J. F\"{o}rster Energy
  Transfer in the Vicinity of Two Metallic Nanospheres (Dimer).
  \emph{Plasmonics} \textbf{2017}, \emph{13}, 873--883\relax
\mciteBstWouldAddEndPuncttrue
\mciteSetBstMidEndSepPunct{\mcitedefaultmidpunct}
{\mcitedefaultendpunct}{\mcitedefaultseppunct}\relax
\EndOfBibitem
\bibitem[Hsu \latin{et~al.}(2017)Hsu, Ding, and Schatz]{Hsu2017}
Hsu,~L.-Y.; Ding,~W.; Schatz,~G.~C. Plasmon-Coupled Resonance Energy Transfer.
  \emph{The Journal of Physical Chemistry Letters} \textbf{2017}, \emph{8},
  2357--2367\relax
\mciteBstWouldAddEndPuncttrue
\mciteSetBstMidEndSepPunct{\mcitedefaultmidpunct}
{\mcitedefaultendpunct}{\mcitedefaultseppunct}\relax
\EndOfBibitem
\bibitem[Sauvan \latin{et~al.}(2013)Sauvan, Hugonin, Maksymov, and
  Lalanne]{sauvan2013theory}
Sauvan,~C.; Hugonin,~J.~P.; Maksymov,~I.~S.; Lalanne,~P. Theory of the
  spontaneous optical emission of nanosize photonic and plasmon resonators.
  \emph{Physical Review Letters} \textbf{2013}, \emph{110}, 237401\relax
\mciteBstWouldAddEndPuncttrue
\mciteSetBstMidEndSepPunct{\mcitedefaultmidpunct}
{\mcitedefaultendpunct}{\mcitedefaultseppunct}\relax
\EndOfBibitem
\bibitem[Doost \latin{et~al.}(2014)Doost, Langbein, and Muljarov]{Muljarov_3D}
Doost,~M.~B.; Langbein,~W.; Muljarov,~E.~A. Resonant-state expansion applied to
  three-dimensional open optical systems. \emph{Physical Review A}
  \textbf{2014}, \emph{90}, 013834\relax
\mciteBstWouldAddEndPuncttrue
\mciteSetBstMidEndSepPunct{\mcitedefaultmidpunct}
{\mcitedefaultendpunct}{\mcitedefaultseppunct}\relax
\EndOfBibitem
\bibitem[Ge \latin{et~al.}(2014)Ge, Kristensen, Young, and Hughes]{Hughes_QNMs}
Ge,~R.-C.; Kristensen,~P.~T.; Young,~J.~F.; Hughes,~S. Quasinormal mode
  approach to modelling light-emission and propagation in nanoplasmonics.
  \emph{New Journal of Physics} \textbf{2014}, \emph{16}, 113048\relax
\mciteBstWouldAddEndPuncttrue
\mciteSetBstMidEndSepPunct{\mcitedefaultmidpunct}
{\mcitedefaultendpunct}{\mcitedefaultseppunct}\relax
\EndOfBibitem
\bibitem[Weiss \latin{et~al.}(2016)Weiss, Mesch, Sch\"{a}ferling, , Giessen,
  Langbein, and Muljarov]{Weiss-Giessen-prl}
Weiss,~T.; Mesch,~M.; Sch\"{a}ferling,~M.; ; Giessen,~H.; Langbein,~W.;
  Muljarov,~E.~A. From Dark to Bright: First-Order Perturbation Theory with
  Analytical Mode Normalization for Plasmonic Nanoantenna Arrays Applied to
  Refractive Index Sensing. \emph{Physical Review Letters} \textbf{2016},
  \emph{116}, 237401\relax
\mciteBstWouldAddEndPuncttrue
\mciteSetBstMidEndSepPunct{\mcitedefaultmidpunct}
{\mcitedefaultendpunct}{\mcitedefaultseppunct}\relax
\EndOfBibitem
\bibitem[Franke \latin{et~al.}(2019)Franke, Hughes, Dezfouli, Kristensen,
  Busch, Knorr, and Richter]{franke2019quantization}
Franke,~S.; Hughes,~S.; Dezfouli,~M.~K.; Kristensen,~P.~T.; Busch,~K.;
  Knorr,~A.; Richter,~M. Quantization of quasinormal modes for open cavities
  and plasmonic cavity quantum electrodynamics. \emph{Physical Review Letters}
  \textbf{2019}, \emph{122}, 213901\relax
\mciteBstWouldAddEndPuncttrue
\mciteSetBstMidEndSepPunct{\mcitedefaultmidpunct}
{\mcitedefaultendpunct}{\mcitedefaultseppunct}\relax
\EndOfBibitem
\bibitem[Alpeggiani \latin{et~al.}(2017)Alpeggiani, Parappurath, Verhagen, and
  Kuipers]{alpeggiani2017quasinormal}
Alpeggiani,~F.; Parappurath,~N.; Verhagen,~E.; Kuipers,~L. Quasinormal-mode
  expansion of the scattering matrix. \emph{Physical Review X} \textbf{2017},
  \emph{7}, 021035\relax
\mciteBstWouldAddEndPuncttrue
\mciteSetBstMidEndSepPunct{\mcitedefaultmidpunct}
{\mcitedefaultendpunct}{\mcitedefaultseppunct}\relax
\EndOfBibitem
\bibitem[Chen \latin{et~al.}(2019)Chen, Bergman, and Sivan]{Chen2019}
Chen,~P.~Y.; Bergman,~D.~J.; Sivan,~Y. Generalizing Normal Mode Expansion of
  Electromagnetic Green's Tensor to Open Systems. \emph{Physical Review
  Applied} \textbf{2019}, \emph{11}\relax
\mciteBstWouldAddEndPuncttrue
\mciteSetBstMidEndSepPunct{\mcitedefaultmidpunct}
{\mcitedefaultendpunct}{\mcitedefaultseppunct}\relax
\EndOfBibitem
\bibitem[Lalanne \latin{et~al.}(2018)Lalanne, Yan, Vynck, Sauvan, and
  Hugonin]{Lalanne2018}
Lalanne,~P.; Yan,~W.; Vynck,~K.; Sauvan,~C.; Hugonin,~J.-P. Light Interaction
  with Photonic and Plasmonic Resonances. \emph{Laser {\&} Photonics Reviews}
  \textbf{2018}, \emph{12}, 1700113\relax
\mciteBstWouldAddEndPuncttrue
\mciteSetBstMidEndSepPunct{\mcitedefaultmidpunct}
{\mcitedefaultendpunct}{\mcitedefaultseppunct}\relax
\EndOfBibitem
\bibitem[Muljarov and Langbein(2016)Muljarov, and
  Langbein]{muljarov2016resonant}
Muljarov,~E.; Langbein,~W. Resonant-state expansion of dispersive open optical
  systems: Creating gold from sand. \emph{Physical Review B} \textbf{2016},
  \emph{93}, 075417\relax
\mciteBstWouldAddEndPuncttrue
\mciteSetBstMidEndSepPunct{\mcitedefaultmidpunct}
{\mcitedefaultendpunct}{\mcitedefaultseppunct}\relax
\EndOfBibitem
\bibitem[Muljarov \latin{et~al.}(2011)Muljarov, Langbein, and
  Zimmermann]{muljarov2011brillouin}
Muljarov,~E.~A.; Langbein,~W.; Zimmermann,~R. Brillouin-Wigner perturbation
  theory in open electromagnetic systems. \emph{Europhysics Letters}
  \textbf{2011}, \emph{92}, 50010\relax
\mciteBstWouldAddEndPuncttrue
\mciteSetBstMidEndSepPunct{\mcitedefaultmidpunct}
{\mcitedefaultendpunct}{\mcitedefaultseppunct}\relax
\EndOfBibitem
\bibitem[Ge \latin{et~al.}(2014)Ge, Kristensen, Young, and
  Hughes]{ge2014quasinormal}
Ge,~R.-C.; Kristensen,~P.~T.; Young,~J.~F.; Hughes,~S. Quasinormal mode
  approach to modelling light-emission and propagation in nanoplasmonics.
  \emph{New Journal of Physics} \textbf{2014}, \emph{16}, 113048\relax
\mciteBstWouldAddEndPuncttrue
\mciteSetBstMidEndSepPunct{\mcitedefaultmidpunct}
{\mcitedefaultendpunct}{\mcitedefaultseppunct}\relax
\EndOfBibitem
\bibitem[Yan \latin{et~al.}(2018)Yan, Faggiani, and Lalanne]{yan2018rigorous}
Yan,~W.; Faggiani,~R.; Lalanne,~P. Rigorous modal analysis of plasmonic
  nanoresonators. \emph{Physical Review B} \textbf{2018}, \emph{97},
  205422\relax
\mciteBstWouldAddEndPuncttrue
\mciteSetBstMidEndSepPunct{\mcitedefaultmidpunct}
{\mcitedefaultendpunct}{\mcitedefaultseppunct}\relax
\EndOfBibitem
\bibitem[Weiss \latin{et~al.}(2017)Weiss, Sch{\"a}ferling, Giessen, Gippius,
  Tikhodeev, Langbein, and Muljarov]{weiss2017analytical}
Weiss,~T.; Sch{\"a}ferling,~M.; Giessen,~H.; Gippius,~N.; Tikhodeev,~S.;
  Langbein,~W.; Muljarov,~E. Analytical normalization of resonant states in
  photonic crystal slabs and periodic arrays of nanoantennas at oblique
  incidence. \emph{Physical Review B} \textbf{2017}, \emph{96}, 045129\relax
\mciteBstWouldAddEndPuncttrue
\mciteSetBstMidEndSepPunct{\mcitedefaultmidpunct}
{\mcitedefaultendpunct}{\mcitedefaultseppunct}\relax
\EndOfBibitem
\bibitem[Bergman and Stroud(1980)Bergman, and Stroud]{Bergman_Stroud_PRB_1980}
Bergman,~D.~J.; Stroud,~D. Theory of resonances in the electromagnetic
  scattering by macroscopic bodies. \emph{Physical Review B} \textbf{1980},
  \emph{22}, 3527--3539\relax
\mciteBstWouldAddEndPuncttrue
\mciteSetBstMidEndSepPunct{\mcitedefaultmidpunct}
{\mcitedefaultendpunct}{\mcitedefaultseppunct}\relax
\EndOfBibitem
\bibitem[Agranovich \latin{et~al.}(1999)Agranovich, Katsenelenbaum, Sivov, and
  Voitovich]{Russians-permittivity-modes-book}
Agranovich,~M.~S.; Katsenelenbaum,~B.~Z.; Sivov,~A.~N.; Voitovich,~N.~N.
  \emph{Generalized method of eigenoscillations in diffraction theory};
  Wiley-VCH, 1999\relax
\mciteBstWouldAddEndPuncttrue
\mciteSetBstMidEndSepPunct{\mcitedefaultmidpunct}
{\mcitedefaultendpunct}{\mcitedefaultseppunct}\relax
\EndOfBibitem
\bibitem[Bergman(1979)]{Bergman_PRB_1979_periodic}
Bergman,~D.~J. Dielectric constant of a two-component granular composite: A
  practical scheme for calculating the pole spectrum. \emph{Physical Review B}
  \textbf{1979}, \emph{19}, 2359--2368\relax
\mciteBstWouldAddEndPuncttrue
\mciteSetBstMidEndSepPunct{\mcitedefaultmidpunct}
{\mcitedefaultendpunct}{\mcitedefaultseppunct}\relax
\EndOfBibitem
\bibitem[Sandu(2013)]{sandu2013eigenmode}
Sandu,~T. Eigenmode decomposition of the near-field enhancement in localized
  surface plasmon resonances of metallic nanoparticles. \emph{Plasmonics}
  \textbf{2013}, \emph{8}, 391--402\relax
\mciteBstWouldAddEndPuncttrue
\mciteSetBstMidEndSepPunct{\mcitedefaultmidpunct}
{\mcitedefaultendpunct}{\mcitedefaultseppunct}\relax
\EndOfBibitem
\bibitem[Bergman(1979)]{bergman1979dielectrica}
Bergman,~D.~J. The dielectric constant of a simple cubic array of identical
  spheres. \emph{J. Phys. C: Solid State Phys.} \textbf{1979}, \emph{12},
  4947\relax
\mciteBstWouldAddEndPuncttrue
\mciteSetBstMidEndSepPunct{\mcitedefaultmidpunct}
{\mcitedefaultendpunct}{\mcitedefaultseppunct}\relax
\EndOfBibitem
\bibitem[Bergman and Dunn(1992)Bergman, and Dunn]{bergman1992bulk}
Bergman,~D.~J.; Dunn,~K.-J. Bulk effective dielectric constant of a composite
  with a periodic microgeometry. \emph{Phys. Rev. B} \textbf{1992}, \emph{45},
  13262--13271\relax
\mciteBstWouldAddEndPuncttrue
\mciteSetBstMidEndSepPunct{\mcitedefaultmidpunct}
{\mcitedefaultendpunct}{\mcitedefaultseppunct}\relax
\EndOfBibitem
\bibitem[Miller \latin{et~al.}(2016)Miller, Polimeridis, Reid, Hsu, DeLacy,
  Joannopoulos, Solja{\v{c}}i{\'{c}}, and Johnson]{Miller2016}
Miller,~O.~D.; Polimeridis,~A.~G.; Reid,~M. T.~H.; Hsu,~C.~W.; DeLacy,~B.~G.;
  Joannopoulos,~J.~D.; Solja{\v{c}}i{\'{c}},~M.; Johnson,~S.~G. Fundamental
  limits to optical response in absorptive systems. \emph{Optics Express}
  \textbf{2016}, \emph{24}, 3329\relax
\mciteBstWouldAddEndPuncttrue
\mciteSetBstMidEndSepPunct{\mcitedefaultmidpunct}
{\mcitedefaultendpunct}{\mcitedefaultseppunct}\relax
\EndOfBibitem
\bibitem[Bergman and Stockman(2003)Bergman, and Stockman]{Bergman2003}
Bergman,~D.~J.; Stockman,~M.~I. Surface Plasmon Amplification by Stimulated
  Emission of Radiation: Quantum Generation of Coherent Surface Plasmons in
  Nanosystems. \emph{Physical Review Letters} \textbf{2003}, \emph{90}\relax
\mciteBstWouldAddEndPuncttrue
\mciteSetBstMidEndSepPunct{\mcitedefaultmidpunct}
{\mcitedefaultendpunct}{\mcitedefaultseppunct}\relax
\EndOfBibitem
\bibitem[Li \latin{et~al.}(2003)Li, Stockman, and Bergman]{li2003self}
Li,~K.; Stockman,~M.~I.; Bergman,~D.~J. Self-Similar Chain of Metal Nanospheres
  as an Efficient Nanolens. \emph{Physical Review Letters} \textbf{2003},
  \emph{91}, 227402\relax
\mciteBstWouldAddEndPuncttrue
\mciteSetBstMidEndSepPunct{\mcitedefaultmidpunct}
{\mcitedefaultendpunct}{\mcitedefaultseppunct}\relax
\EndOfBibitem
\bibitem[Stockman \latin{et~al.}(2004)Stockman, Bergman, Anceau, Brasselet, and
  Zyss]{Stockman:2004ey}
Stockman,~M.; Bergman,~D.; Anceau,~C.; Brasselet,~S.; Zyss,~J. {Enhanced
  Second-Harmonic Generation by Metal Surfaces with Nanoscale Roughness:
  Nanoscale Dephasing, Depolarization, and Correlations}. \emph{Physical Review
  Letters} \textbf{2004}, \emph{92}, 057402\relax
\mciteBstWouldAddEndPuncttrue
\mciteSetBstMidEndSepPunct{\mcitedefaultmidpunct}
{\mcitedefaultendpunct}{\mcitedefaultseppunct}\relax
\EndOfBibitem
\bibitem[Stockman \latin{et~al.}(2002)Stockman, Faleev, and
  Bergman]{Stockman2002}
Stockman,~M.~I.; Faleev,~S.~V.; Bergman,~D.~J. Coherent Control of Femtosecond
  Energy Localization in Nanosystems. \emph{Physical Review Letters}
  \textbf{2002}, \emph{88}\relax
\mciteBstWouldAddEndPuncttrue
\mciteSetBstMidEndSepPunct{\mcitedefaultmidpunct}
{\mcitedefaultendpunct}{\mcitedefaultseppunct}\relax
\EndOfBibitem
\bibitem[Li \latin{et~al.}(2005)Li, Stockman, and Bergman]{Li2005}
Li,~K.; Stockman,~M.~I.; Bergman,~D.~J. Enhanced second harmonic generation in
  a self-similar chain of metal nanospheres. \emph{Physical Review B}
  \textbf{2005}, \emph{72}\relax
\mciteBstWouldAddEndPuncttrue
\mciteSetBstMidEndSepPunct{\mcitedefaultmidpunct}
{\mcitedefaultendpunct}{\mcitedefaultseppunct}\relax
\EndOfBibitem
\bibitem[Reddy \latin{et~al.}(2017)Reddy, Chen, Fern\'andez-Dom\'inguez, and
  Sivan]{Reddy_Sivan_SHG_BCs}
Reddy,~K.~N.; Chen,~P.~Y.; Fern\'andez-Dom\'inguez,~A.~I.; Sivan,~Y. Revisiting
  the boundary conditions for second-harmonic generation at metal-dielectric
  interfaces. \emph{Journal of the Optical Society of America B} \textbf{2017},
  \emph{34}, 1824\relax
\mciteBstWouldAddEndPuncttrue
\mciteSetBstMidEndSepPunct{\mcitedefaultmidpunct}
{\mcitedefaultendpunct}{\mcitedefaultseppunct}\relax
\EndOfBibitem
\bibitem[Farhi and Bergman(2014)Farhi, and Bergman]{Farhi_Bergman_PRA_2014}
Farhi,~A.; Bergman,~D. Analysis of a Veselago lens in the quasistatic regime.
  \emph{Physical Review A} \textbf{2014}, \emph{90}, 013806\relax
\mciteBstWouldAddEndPuncttrue
\mciteSetBstMidEndSepPunct{\mcitedefaultmidpunct}
{\mcitedefaultendpunct}{\mcitedefaultseppunct}\relax
\EndOfBibitem
\bibitem[Schnitzer \latin{et~al.}(2016)Schnitzer, Giannini, Maier, and
  Craster]{Schnitzer_2016}
Schnitzer,~O.; Giannini,~V.; Maier,~S.~A.; Craster,~R.~V. Surface plasmon
  resonances of arbitrarily shaped nanometallic structures in the
  small-screening-length limit. \emph{Proceedings of the Royal Society A}
  \textbf{2016}, \emph{472}, 20160258\relax
\mciteBstWouldAddEndPuncttrue
\mciteSetBstMidEndSepPunct{\mcitedefaultmidpunct}
{\mcitedefaultendpunct}{\mcitedefaultseppunct}\relax
\EndOfBibitem
\bibitem[Schnitzer(2019)]{Schnitzer_2019}
Schnitzer,~O. Geometric quantization of localized surface plasmons. \emph{IMA
  Journal of Applied Mathematics} \textbf{2019}, \emph{84}\relax
\mciteBstWouldAddEndPuncttrue
\mciteSetBstMidEndSepPunct{\mcitedefaultmidpunct}
{\mcitedefaultendpunct}{\mcitedefaultseppunct}\relax
\EndOfBibitem
\bibitem[Farhi and Bergman(2016)Farhi, and Bergman]{Farhi_Bergman_PRA_2016}
Farhi,~A.; Bergman,~D. Electromagnetic eigenstates and the field of an
  oscillating point electric dipole in a flat-slab composite structure.
  \emph{Physical Review A} \textbf{2016}, \emph{93}, 063844\relax
\mciteBstWouldAddEndPuncttrue
\mciteSetBstMidEndSepPunct{\mcitedefaultmidpunct}
{\mcitedefaultendpunct}{\mcitedefaultseppunct}\relax
\EndOfBibitem
\bibitem[Chen and Sivan(2017)Chen, and Sivan]{Parry-wire-principal_value}
Chen,~P.~Y.; Sivan,~Y. Robust Location of Optical Fiber Modes via the Argument
  Principle Method. \emph{Computer Physics Communications} \textbf{2017},
  \emph{214}, 105--116\relax
\mciteBstWouldAddEndPuncttrue
\mciteSetBstMidEndSepPunct{\mcitedefaultmidpunct}
{\mcitedefaultendpunct}{\mcitedefaultseppunct}\relax
\EndOfBibitem
\bibitem[Hoehne \latin{et~al.}(2018)Hoehne, Zschiedrich, Haghighi, Lott, and
  Burger]{hoehne2018validation}
Hoehne,~T.; Zschiedrich,~L.; Haghighi,~N.; Lott,~J.~A.; Burger,~S. Validation
  of quasi-normal modes and of constant-flux modes for computing fundamental
  resonances of VCSELs. Semiconductor Lasers and Laser Dynamics VIII. 2018; p
  106821U\relax
\mciteBstWouldAddEndPuncttrue
\mciteSetBstMidEndSepPunct{\mcitedefaultmidpunct}
{\mcitedefaultendpunct}{\mcitedefaultseppunct}\relax
\EndOfBibitem
\bibitem[Ge \latin{et~al.}(2010)Ge, Chong, and Stone]{ge2010steady}
Ge,~L.; Chong,~Y.; Stone,~A.~D. Steady-state ab initio laser theory:
  generalizations and analytic results. \emph{Physical Review A} \textbf{2010},
  \emph{82}, 063824\relax
\mciteBstWouldAddEndPuncttrue
\mciteSetBstMidEndSepPunct{\mcitedefaultmidpunct}
{\mcitedefaultendpunct}{\mcitedefaultseppunct}\relax
\EndOfBibitem
\bibitem[Huheey(1972)]{LCAO-book}
Huheey,~J. \emph{Inorganic Chemistry:Principles of Structure and Reactivity};
  Prentice Hall; 4 edition, 1972\relax
\mciteBstWouldAddEndPuncttrue
\mciteSetBstMidEndSepPunct{\mcitedefaultmidpunct}
{\mcitedefaultendpunct}{\mcitedefaultseppunct}\relax
\EndOfBibitem
\bibitem[Gersten and Nitzan(1985)Gersten, and Nitzan]{Gersten_Nitzan_review}
Gersten,~J.~I.; Nitzan,~A. Photophysics and photochemistry near surfaces and
  small particles. \emph{Surface Science} \textbf{1985}, \emph{158}, 165\relax
\mciteBstWouldAddEndPuncttrue
\mciteSetBstMidEndSepPunct{\mcitedefaultmidpunct}
{\mcitedefaultendpunct}{\mcitedefaultseppunct}\relax
\EndOfBibitem
\bibitem[Liver \latin{et~al.}(1985)Liver, Nitzan, and
  Freed]{Nitzan_hybridization}
Liver,~N.; Nitzan,~A.; Freed,~K.~F. Radiative and nonradiative decay rates of
  molecules adsorbed on clusters of small dielectric particles. \emph{Journal
  of Chemical Physics} \textbf{1985}, \emph{82}, 3831\relax
\mciteBstWouldAddEndPuncttrue
\mciteSetBstMidEndSepPunct{\mcitedefaultmidpunct}
{\mcitedefaultendpunct}{\mcitedefaultseppunct}\relax
\EndOfBibitem
\bibitem[Liver \latin{et~al.}(1984)Liver, Nitzan, and
  Gersten]{Gersten_Nitzan_hybridization}
Liver,~N.; Nitzan,~A.; Gersten,~J. Local fields in cavity sites of rough
  dielectric surfaces. \emph{Chemical Physics Letters} \textbf{1984},
  \emph{111}, 449--454\relax
\mciteBstWouldAddEndPuncttrue
\mciteSetBstMidEndSepPunct{\mcitedefaultmidpunct}
{\mcitedefaultendpunct}{\mcitedefaultseppunct}\relax
\EndOfBibitem
\bibitem[Wang \latin{et~al.}(1993)Wang, Zhang, Yu, and
  Harmon]{wang1993multiple}
Wang,~X.; Zhang,~X.-G.; Yu,~Q.; Harmon,~B. Multiple-scattering theory for
  electromagnetic waves. \emph{Physical Review B} \textbf{1993}, \emph{47},
  4161\relax
\mciteBstWouldAddEndPuncttrue
\mciteSetBstMidEndSepPunct{\mcitedefaultmidpunct}
{\mcitedefaultendpunct}{\mcitedefaultseppunct}\relax
\EndOfBibitem
\bibitem[Chen \latin{et~al.}(2012)Chen, Byrne, Asatryan, Botten, Dossou, Tuniz,
  McPhedran, De~Sterke, Poulton, and Steel]{chen2012plane}
Chen,~P.; Byrne,~M.; Asatryan,~A.; Botten,~L.; Dossou,~K.; Tuniz,~A.;
  McPhedran,~R.; De~Sterke,~C.; Poulton,~C.; Steel,~M. Plane-wave scattering by
  a photonic crystal slab: Multipole modal formulation and accuracy.
  \emph{Waves in Random and Complex Media} \textbf{2012}, \emph{22},
  531--570\relax
\mciteBstWouldAddEndPuncttrue
\mciteSetBstMidEndSepPunct{\mcitedefaultmidpunct}
{\mcitedefaultendpunct}{\mcitedefaultseppunct}\relax
\EndOfBibitem
\bibitem[Prodan \latin{et~al.}(2003)Prodan, Radloff, Halas, and
  Nordlander]{hybridization_science_2003}
Prodan,~E.; Radloff,~C.; Halas,~N.~J.; Nordlander,~P. A Hybridization Model for
  the Plasmon Response of Complex Nanostructures. \emph{Science} \textbf{2003},
  \emph{302}, 419\relax
\mciteBstWouldAddEndPuncttrue
\mciteSetBstMidEndSepPunct{\mcitedefaultmidpunct}
{\mcitedefaultendpunct}{\mcitedefaultseppunct}\relax
\EndOfBibitem
\bibitem[Nordlander \latin{et~al.}(2004)Nordlander, Oubre, Prodan, Li, and
  Stockman]{hybridization_NL_2004}
Nordlander,~P.; Oubre,~C.; Prodan,~E.; Li,~K.; Stockman,~M.~I. Plasmon
  Hybridization in nanoparticle dimers. \emph{Science} \textbf{2004}, \emph{4},
  899--903\relax
\mciteBstWouldAddEndPuncttrue
\mciteSetBstMidEndSepPunct{\mcitedefaultmidpunct}
{\mcitedefaultendpunct}{\mcitedefaultseppunct}\relax
\EndOfBibitem
\bibitem[Onl()]{OnlineCodes}
Code available on www.umons.ac.be/nanophot\relax
\mciteBstWouldAddEndPuncttrue
\mciteSetBstMidEndSepPunct{\mcitedefaultmidpunct}
{\mcitedefaultendpunct}{\mcitedefaultseppunct}\relax
\EndOfBibitem
\bibitem[Rivera \latin{et~al.}(2016)Rivera, Kaminer, Zhen, Joannopoulos, and
  Solja{\v{c}}i{\'{c}}]{Rivera2016}
Rivera,~N.; Kaminer,~I.; Zhen,~B.; Joannopoulos,~J.~D.;
  Solja{\v{c}}i{\'{c}},~M. Shrinking light to allow forbidden transitions on
  the atomic scale. \emph{Science} \textbf{2016}, \emph{353}, 263--269\relax
\mciteBstWouldAddEndPuncttrue
\mciteSetBstMidEndSepPunct{\mcitedefaultmidpunct}
{\mcitedefaultendpunct}{\mcitedefaultseppunct}\relax
\EndOfBibitem
\bibitem[Castani\'e \latin{et~al.}(2010)Castani\'e, Boffety, and
  Carminati]{Carminati_nonlocal_local}
Castani\'e,~E.; Boffety,~M.; Carminati,~R. Fluorescence quenching by a metal
  nanoparticle in the extreme near-field regime. \emph{Optics Letters}
  \textbf{2010}, \emph{35}, 291\relax
\mciteBstWouldAddEndPuncttrue
\mciteSetBstMidEndSepPunct{\mcitedefaultmidpunct}
{\mcitedefaultendpunct}{\mcitedefaultseppunct}\relax
\EndOfBibitem
\bibitem[Aubry \latin{et~al.}(2011)Aubry, Lei, Maier, and
  Pendry]{Alex_hybridization}
Aubry,~A.; Lei,~D.; Maier,~S.~A.; Pendry,~J.~B. Plasmonic Hybridization between
  Nanowires and a Metallic Surface: A Transformation Optics Approach. \emph{ACS
  Nano} \textbf{2011}, \emph{5}, 3293--3308\relax
\mciteBstWouldAddEndPuncttrue
\mciteSetBstMidEndSepPunct{\mcitedefaultmidpunct}
{\mcitedefaultendpunct}{\mcitedefaultseppunct}\relax
\EndOfBibitem
\bibitem[Giannini \latin{et~al.}(2011)Giannini, Francescato, Amrania, Phillips,
  and Maier]{Giannini2011}
Giannini,~V.; Francescato,~Y.; Amrania,~H.; Phillips,~C.~C.; Maier,~S.~A. Fano
  Resonances in Nanoscale Plasmonic Systems: A Parameter-Free Modeling
  Approach. \emph{Nano Letters} \textbf{2011}, \emph{11}, 2835--2840\relax
\mciteBstWouldAddEndPuncttrue
\mciteSetBstMidEndSepPunct{\mcitedefaultmidpunct}
{\mcitedefaultendpunct}{\mcitedefaultseppunct}\relax
\EndOfBibitem
\bibitem[Zhang \latin{et~al.}(2008)Zhang, Genov, Wang, Liu, and
  Zhang]{Zhang2008}
Zhang,~S.; Genov,~D.~A.; Wang,~Y.; Liu,~M.; Zhang,~X. Plasmon-Induced
  Transparency in Metamaterials. \emph{Physical Review Letters} \textbf{2008},
  \emph{101}\relax
\mciteBstWouldAddEndPuncttrue
\mciteSetBstMidEndSepPunct{\mcitedefaultmidpunct}
{\mcitedefaultendpunct}{\mcitedefaultseppunct}\relax
\EndOfBibitem
\bibitem[Hentschel \latin{et~al.}(2011)Hentschel, Dregely, Vogelgesang,
  Giessen, and Liu]{Giessen_Liu_oligomers}
Hentschel,~M.; Dregely,~D.; Vogelgesang,~R.; Giessen,~H.; Liu,~N. Plasmonic
  Oligomers: The Role of Individual Particles in Collective Behavior. \emph{ACS
  Nano} \textbf{2011}, \emph{5}, 2042--2050\relax
\mciteBstWouldAddEndPuncttrue
\mciteSetBstMidEndSepPunct{\mcitedefaultmidpunct}
{\mcitedefaultendpunct}{\mcitedefaultseppunct}\relax
\EndOfBibitem
\bibitem[Rivera \latin{et~al.}(2019)Rivera, Wong, Solja\ifmmode \check{c}\else
  \v{c}\fi{}i\ifmmode~\acute{c}\else \'{c}\fi{}, and Kaminer]{Rivera2019}
Rivera,~N.; Wong,~L.~J.; Solja\ifmmode \check{c}\else
  \v{c}\fi{}i\ifmmode~\acute{c}\else \'{c}\fi{},~M.; Kaminer,~I. Ultrafast
  Multiharmonic Plasmon Generation by Optically Dressed Electrons.
  \emph{Physical Review Letters} \textbf{2019}, \emph{122}, 053901\relax
\mciteBstWouldAddEndPuncttrue
\mciteSetBstMidEndSepPunct{\mcitedefaultmidpunct}
{\mcitedefaultendpunct}{\mcitedefaultseppunct}\relax
\EndOfBibitem
\bibitem[Rosolen \latin{et~al.}(2018)Rosolen, Wong, Rivera, Maes,
  Solja{\v{c}}i{\'{c}}, and Kaminer]{Rosolen2018}
Rosolen,~G.; Wong,~L.~J.; Rivera,~N.; Maes,~B.; Solja{\v{c}}i{\'{c}},~M.;
  Kaminer,~I. Metasurface-based multi-harmonic free-electron light source.
  \emph{Light: Science {\&} Applications} \textbf{2018}, \emph{7}\relax
\mciteBstWouldAddEndPuncttrue
\mciteSetBstMidEndSepPunct{\mcitedefaultmidpunct}
{\mcitedefaultendpunct}{\mcitedefaultseppunct}\relax
\EndOfBibitem
\bibitem[Guo and Jacob(2013)Guo, and Jacob]{Guo_Thermal_MTM}
Guo,~Y.; Jacob,~Z. Thermal hyperbolic metamaterials. \emph{Optics Express}
  \textbf{2013}, \emph{21}, 15014--15019\relax
\mciteBstWouldAddEndPuncttrue
\mciteSetBstMidEndSepPunct{\mcitedefaultmidpunct}
{\mcitedefaultendpunct}{\mcitedefaultseppunct}\relax
\EndOfBibitem
\bibitem[Guo \latin{et~al.}(2014)Guo, Molesky, Hu, Cortes, and
  Jacob]{Guo_Thermal_MTM_TPV}
Guo,~Y.; Molesky,~S.; Hu,~H.; Cortes,~C.~L.; Jacob,~Z. Thermal excitation of
  plasmons for near-field thermophotovoltaics. \emph{Applied Physics Letters}
  \textbf{2014}, \emph{105}, 073903\relax
\mciteBstWouldAddEndPuncttrue
\mciteSetBstMidEndSepPunct{\mcitedefaultmidpunct}
{\mcitedefaultendpunct}{\mcitedefaultseppunct}\relax
\EndOfBibitem
\bibitem[Francoeur \latin{et~al.}(2009)Francoeur, Meng\"{u}{\c{c}}, and
  Vaillon]{Francoeur2009}
Francoeur,~M.; Meng\"{u}{\c{c}},~M.~P.; Vaillon,~R. Solution of near-field
  thermal radiation in one-dimensional layered media using dyadic Green's
  functions and the scattering matrix method. \emph{Journal of Quantitative
  Spectroscopy and Radiative Transfer} \textbf{2009}, \emph{110},
  2002--2018\relax
\mciteBstWouldAddEndPuncttrue
\mciteSetBstMidEndSepPunct{\mcitedefaultmidpunct}
{\mcitedefaultendpunct}{\mcitedefaultseppunct}\relax
\EndOfBibitem
\bibitem[Czapla and Narayanaswamy(2017)Czapla, and Narayanaswamy]{Czapla2017}
Czapla,~B.; Narayanaswamy,~A. Near-field thermal radiative transfer between two
  coated spheres. \emph{Physical Review B} \textbf{2017}, \emph{96}\relax
\mciteBstWouldAddEndPuncttrue
\mciteSetBstMidEndSepPunct{\mcitedefaultmidpunct}
{\mcitedefaultendpunct}{\mcitedefaultseppunct}\relax
\EndOfBibitem
\bibitem[Luo \latin{et~al.}(2014)Luo, Zhao, and Pendry]{Luo2014}
Luo,~Y.; Zhao,~R.; Pendry,~J.~B. van der Waals interactions at the nanoscale:
  The effects of nonlocality. \emph{Proceedings of the National Academy of
  Sciences} \textbf{2014}, \emph{111}, 18422--18427\relax
\mciteBstWouldAddEndPuncttrue
\mciteSetBstMidEndSepPunct{\mcitedefaultmidpunct}
{\mcitedefaultendpunct}{\mcitedefaultseppunct}\relax
\EndOfBibitem
\bibitem[Mosk \latin{et~al.}(2012)Mosk, Lagendijk, Lerosey, and
  Fink]{Mosk_review_NP}
Mosk,~A.~P.; Lagendijk,~A.; Lerosey,~G.; Fink,~M. Controlling waves in space
  and time for imaging and focusing in complex media. \emph{Nature Photonics}
  \textbf{2012}, \emph{6}, 283\relax
\mciteBstWouldAddEndPuncttrue
\mciteSetBstMidEndSepPunct{\mcitedefaultmidpunct}
{\mcitedefaultendpunct}{\mcitedefaultseppunct}\relax
\EndOfBibitem
\end{mcitethebibliography}

\end{document}